# Wavefront shaping for optimized many-mode Kerr beam self-cleaning in graded-index multimode fiber

E. Deliancourt[1], M. Fabert[1], A. Tonello[1], K. Krupa[2],
A. Desfarges-Berthelemot[1], V. Kermene[1], G. Millot[2],
A. Barthélémy[1], S. Wabnitz[3], and V. Couderc [1,*]

[1] *Université de Limoges, XLIM, UMR CNRS 7252, 123 Av. A. Thomas, 87060 Limoges, France*
[2] *Université Bourgogne Franche-Comté, ICB, UMR CNRS 6303, 9 Av. A. Savary, 21078 Dijon, France*
[3] *DIET, Sapienza University of Rome, Via Eudossiana 18, 00184 Rome, Italy*
*\*vincent.couderc@xlim.fr*

**Abstract:** We report experimental results, showing that the Kerr beam self-cleaning of many low-order modes in a graded-index multimode fiber can be controlled thanks to optimized wavefront shaping of the coherent excitation beam. Adaptive profiling of the transverse input phase was utilized for channeling the launched power towards a specific low-order fiber mode, by exploiting nonlinear coupling among all guided modes. Experiments were carried out with 7 ps pulses at 1064 nm injected in a five meters long multimode fiber operating in the normal dispersion regime. Optimized Kerr beam self-cleaning of five different LP modes is reported, with a power threshold that increases with the mode order.

## 1. Introduction

In the context of research on light wave propagation through complex systems, the availability of liquid crystal-based spatial light modulators with a large number of pixels has permitted breakthrough experiments. Among them, we can mention the pioneering works involving the control of light propagation through scattering media. In that context, it was shown that input wavefront shaping of a light beam may serve for output beam pattern control [1], for transmission matrix measurement and image recovery [2,3], for the transmission of ultrashort pulses [4], for the maximization of power transmission [5], for space-time control [6], etc. All of these advances were based on the principle that the field at the output of the disordered medium results from an interference of many waves, corresponding to the different paths which light can follow along its propagation through the bulk medium. Special phase adjustments of the different degrees of freedom in the input transverse plane (pixels) give a way to control the output state of interference, in both spatial and temporal dimensions. Almost all published works deal with laser light propagation through linear scattering media, which is the situation that is most frequently encountered. Only a limited number of works has appeared, where Kerr nonlinearity was involved. One of such examples is the recent work by Frostig *et al.* [7], which reported adaptive beam focusing through a scattering layer, followed by a layer with Kerr nonlinearity.

Light propagation through multimode optical fibers (MMF) shares several analogies with the problem of light scattering in complex heterogeneous media [8]. In fact, along the course of beam propagation in a MMF, longitudinal mode coupling caused by imperfections, stress or bending variations plays a role which is analogous to that of scattering from refractive index disorder in a bulk medium. The transmitted light results in both cases from the interference of many fields with a complex amplitude and phase transverse distribution, leading to a speckled intensity pattern with a seemingly random spatial distribution. On the other hand, there are also some major differences between the case of MMFs and random bulk media. In short segments of multimode fiber, one may neglect backscattering radiation (propagation models are unidirectional), whereas in disordered media backscattered light is commonly taken into account. Moreover, in the case of standard waveguides, the appropriate basis for the output field expansion is discrete and bounded (the MMF eigenmodes). Techniques which were previously developed for shaping both in space and in time a wave transmitted through disordered media by means of input wavefront structuration, have been subsequently successfully adapted to the case of MMFs. The delivery of focused beams or of a beam with a given specific pattern at the output of a MMF was demonstrated [9,10]. In addition, the imaging of biological samples through MMF [11] was achieved. Again, most of experiments reported so far with MMFs dealt with the linear propagation regime, the case of an Yb-doped MMF amplifier operated in the saturated gain regime being a noticeable exception [12]. However, since MMFs tightly confine light power over long propagation distances, the onset of nonlinear effects may be easily reached.

Nonlinear coupling among guided modes in a MMF, for instance through four-wave mixing, produces nonlinear scattering effects which adds up to the already existing linear coupling mechanisms. Thus, MMFs offer a convenient platform for the study of many-mode nonlinear interactions, and their space-time dynamics. That is partly one of the reasons why there has been a renewed interest in studying the multimode nonlinear propagation regime in optical fibers. Striking experiments on multimode optical solitons [13], on geometric parametric instability [14] and beam self-cleaning in GRIN MMFs [15, 16, 17], and on supercontinuum generation [18, 19], have been reported in very recent years. In almost all of the publications related to this topic, a standard circular laser beam with a Gaussian cross section and linear state of polarization was used at the input of the MMF. Only the degrees of freedom provided by the input diameter and by the incidence angle were modified, in order to control the efficiency of the nonlinear effects. It is not until very recently that an input beam with a tailored shape has been used for enhancing or mitigating frequency conversion resulting from either modal-FWM or stimulated Raman scattering (SRS) in MMFs [20, 21].

In this work we experimentally study, we believe for the first time, the application of wavefront shaping for the control of the transverse output modal distribution resulting from nonlinear propagation in MMFs. More particularly, wavefront shaping permits us to exploit all of the spatial degrees of freedom of the input beam to control Kerr beam self-cleaning processes in a GRIN MMF. Adaptive wavefront profiling of the excitation pulse leads to optimized (via minimization of an appropriate cost function) self-cleaning of the output radiation on many different low-order mode (LOM) structures, as the input laser power was progressively increased.

## 2. Experimental set-up and procedure

In the experiments, we have used as the light source an Nd:YVO4 ultrafast laser, delivering 7 ps pulses at 1064 nm and 1 MHz repetition rate (Sirius Spark Laser), with a pulse energy of up to 1 µJ. The laser beam was first expanded with a magnifying telescope, to cover the reflective surface of a deformable mirror (DM) with 140 actuators (see Fig.1). The DM had a continuous metallic membrane with a square array of actuators (~12x12). The beam illuminated the mirror with an almost normal incidence angle. The reflected beam was imaged on the input facet of a fiber after a strong down scaling by means of a second telescope. A half-wave plate (HWP) permitted to vary the linear polarization orientation of the beam. The shaped beam was launched into a piece of GRIN-MMF of about 5 m in length. The fiber was freely wounded and laid on the optical table making rings of about 20 cm in diameter. At the fiber output, a short focal length lens imaged the near field intensity on a camera, after bandpass filtering (10 nm at 1064 nm) and selection of a single linear polarization state. The beam was split in order to simultaneously display the fiber output far field on a second camera. An optical spectrum analyzer (OSA) completed the output pulsed beam diagnosis tools. The near-field image taken by the camera was displayed on a screen for preliminary adjustments and for observations.

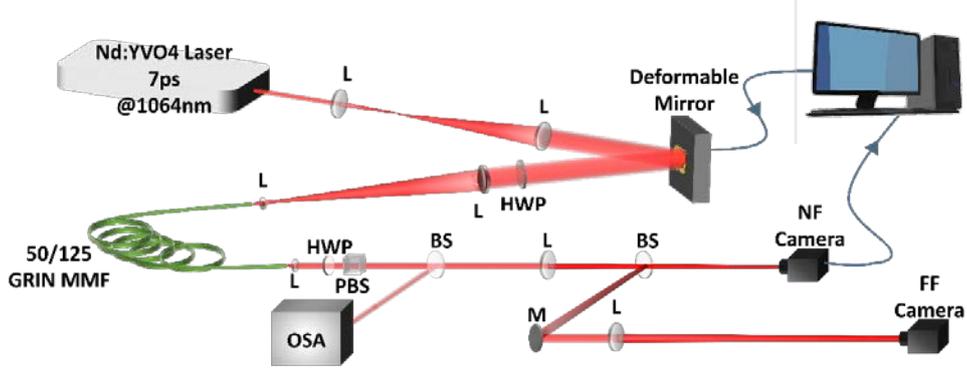

Fig. 1. Experimental set-up including a laser source, lenses (L) for magnification and imaging, a 5 m long GRIN-MMF, a deformable mirror with 140 actuators, waveplates (HWP) for polarization state orientation, non-polarizing (BS) and polarizing beam splitters (PBS), optical spectrum analyzer (OSA), cameras and a lap top.

At the same time, the near-field image was also sent to a laptop, for computing parameters which served for the optimization of the input wavefront. In a preliminary step of adjustment, the deformable mirror was set to a uniform state. Its mechanical orientation as well as the position of the lens for injection in the MMF were finely tuned to obtain efficient coupling into the fiber, and to observe self-cleaning on the fundamental mode, as the launched beam power was progressively raised. Optimization of the input wavefront was based on the intensity correlation between the experimental output image $I_{exp}$ and the target output mode pattern $I_{tar}$, in a way to minimize the cost function [22]

$$C = 1 - \langle (I_{exp} - \bar{I}_{exp})(I_{tar} - \bar{I}_{tar}) \rangle / \sqrt{\langle (I_{exp} - \bar{I}_{exp})^2 \rangle \langle (I_{tar} - \bar{I}_{tar})^2 \rangle} . \qquad (1)$$

where $\bar{I}_h = \langle I_h \rangle$, and $\langle \cdot \rangle$ denotes integration over the transverse domain. Our approach was based on a sequential optimization algorithm that was derived from the seminal papers of Vellekoop [1, 23]. The pixels of the mirror were successively randomly selected by groups of 70, 40, 30, 20 and 10 elements. For each group size, the pixels to optimize were randomly chosen, and the cost function was evaluated while the phase of the pixels was cycled between 0 and $2\pi$. The phase value which minimized the parameter C was then set on the relevant mirror elements, and the random choice of pixels was repeated 70, 140, 210, 280 and 350 times, respectively, for each different group of the previously mentioned elements.

## 3. Results and discussion

In a preliminary step, and with the DM in a uniform flat setting, as previously mentioned we obtained Kerr beam self-cleaning on the fundamental mode (or $LP_{01}$ mode) for the launched peak power of 45 kW. We set this input power level, in order to maintain sufficient nonlinearity to trigger self-cleaning on different low-order modes, after that spatial structuration of the input beam (with its associated loss of coupling efficiency) was introduced. We verified that for the fiber length used, and for the maximum power injected into the MMF, self-stimulated Raman scattering could be neglected. The laser power level was then kept fixed for the following experiments. Upon hand adjusting the laser beam injection into the MMF, we obtained an output intensity pattern as shown on Fig.2-a. The pattern of Fig.2-a represents a typical self-cleaned beam, whereby the Kerr effect leads to a central sharp spot, surrounded by a speckled background of much lower intensity. Similar features appear in the output far field. The coupling conditions, *i.e.*, the position of the focusing lens and the position of the fiber input facet, were kept fixed in all of the following experiments. Only the structuration of the DM was varied in the next steps.

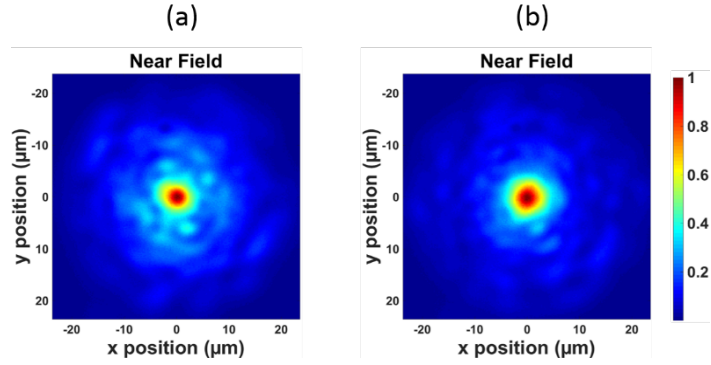

Fig. 2. (a) Standard Kerr beam self-cleaning obtained with a flat wavefront input beam and observed at the fiber output with a Peak power of 45 kW. (b) Kerr beam self-cleaning on the fundamental mode of the fiber after adaptive input wavefront shaping (Peak power 40 kW).

At first, we started with an optimization of the wavefront, with the $LP_{01}$ mode as a target. Even though self-cleaning was already present in the initial state, the transverse structuration of the input wave led to a sensible improvement of the self-cleaned beam quality. The cost function C to minimize dropped from 0.19 down to 0.075 after convergence, and the intensity correlation between the experimental profile and the theoretical one grew from 0.66 to 0.75. The main central spot was slightly broadened, while the surrounding background shrank and became smoother, as illustrated on the recordings of Fig.2-b.

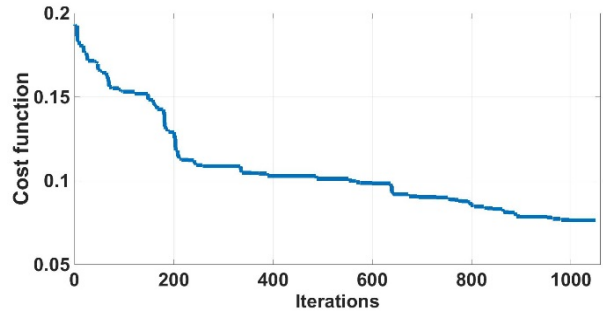

Fig. 3. Evolution of the cost function according to the iterations of the optimization process for Kerr self-cleaning on the fundamental mode of the GRIN MMF.

It takes 1050 iterations of our algorithm, in order to converge towards a steady state, a process which takes ~4 minutes (see Fig.3). At the same time, the coupled power into the MMF decreased by 11%. The DM was reset to a flat state before making a new optimization round, this time with the $LP_{11}$ mode as a target. Tailoring of the input yielded Kerr self-cleaning on the $LP_{11}$ structure, as expected (see Fig.4). This result is similar to that reported in [24], except that in that reference the self-cleaning of the $LP_{11}$ mode was obtained with a lengthy and difficult to exactly reproduce manual procedure, i.e., with a tricky adjustment of tilted launching conditions. The occurrence of a similar shape and orientation of the intensity pattern both in the near field and far field attests that the phase distribution of the output field is in fact the one expected for an $LP_{11}$ eigenmode (see the Appendix for details). This proves that the output field contained a major contribution of the target mode. However, note that the optimization of the DM surface was driven by the near field intensity profile only.

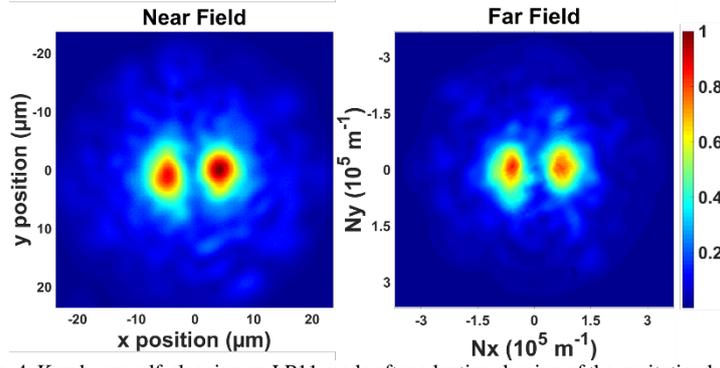

Fig. 4. Kerr beam self-cleaning on LP11 mode after adaptive shaping of the excitation beam wavefront (Peak power 34.5 kW). The occurrence of similar patterns in the near and far fields demonstrate the major content of the LP11 in the output modal population.

Our aim was also to generate higher-order modes. Hence, we successively targeted $LP_{02}$, $LP_{21}$, $LP_{12}$, and finally $LP_{22}$ modes. In all cases, once the optimization of the input wavefront reached a steady state, we obtained output intensity pictures closely resembling the pattern of the target mode, although with a lower quality by comparison to the previously generated $LP_{11}$ mode. Similar to the case of the $LP_{11}$ mode, the recorded near and far field patterns exhibited the same intensity profile, thus demonstrating the major content of the sought mode in the output field. Typical recordings are shown in Fig.5.

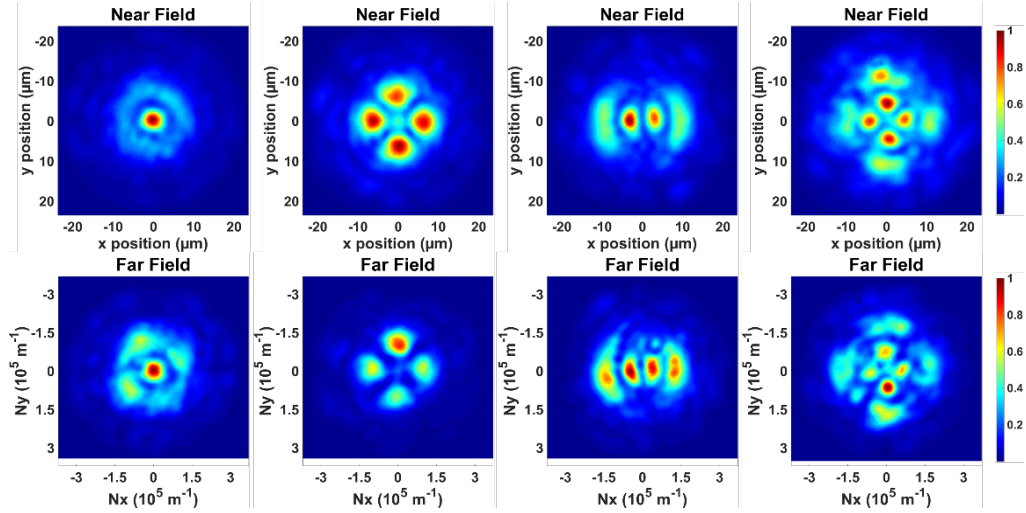

Fig. 5. Near field (top row) and far field (bottom row) images recorded after optimization of the input wavefront, in order to obtain a desired mode profile at the MMF output facet ($LP_{02}$, $LP_{21}$, $LP_{12}$ and $LP_{22}$ from left to right, respectively) at relatively high peak powers (31 kW, 31 kW, 29 kW and 29 kW, respectively)

Intensity correlations between the experimental profile and the theoretical mode profile reached values in the range of 0.75-0.82, for all of the studied modes. As far as the fiber power coupling loss introduced by the beam structuring is concerned, its value varied from one mode to another (11% for the $LP_{11}$ mode, 28% for the $LP_{01}$, 15% for the $LP_{02}$, 16% for the $LP_{21}$, 22% for the $LP_{12}$, 21% for the $LP_{22}$) in a given experimental setting. A series of experiments was carried out for various coupled power levels and various targeted modes, with the purpose of achieving optimized self-beam cleaning. We identified that the power threshold $P_{th}$ to reach Kerr self-cleaning increased with the order of the desired mode. Such threshold was $P_{th}$~6.8 kW for the $LP_{01}$, $P_{th}$ ~16.5 kW for the $LP_{11}$ and $P_{th}$ ~20.5 kW for the $LP_{21}$ mode, respectively.

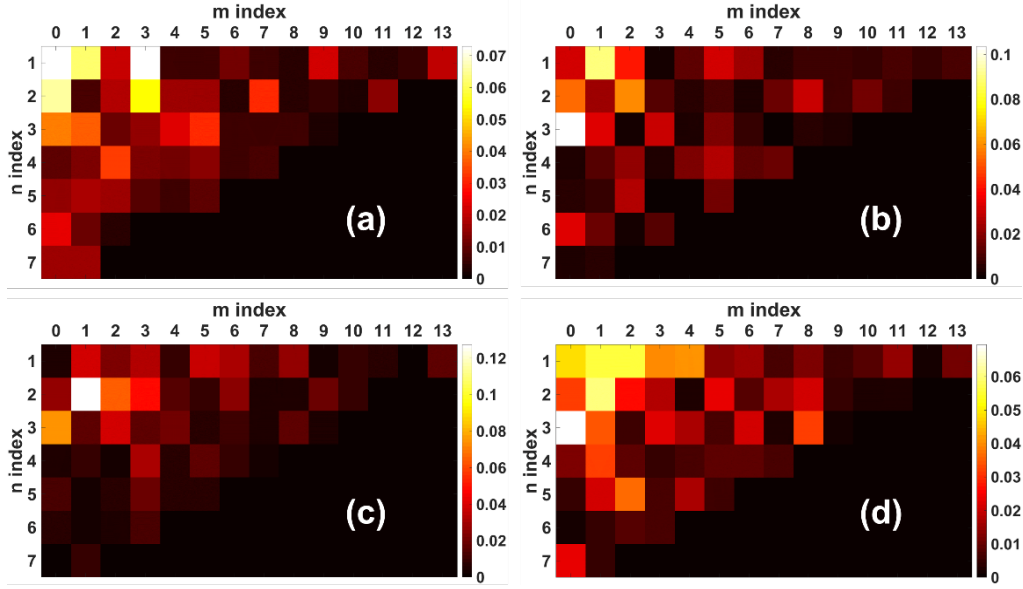

Fig. 6. Mode power distribution of the launched fields, in order to obtain self-cleaning on the $LP_{01}$ (a), $LP_{11}$ (b), $LP_{12}$ (c), and $LP_{21}$ (d) modes, respectively. Charts were computed from projections of the optimized wavefront profiles on the fiber mode basis. Color bars indicate the percentage of input mode power.

Once that the optimized wavefront profile is known, assuming that the demagnifying imaging between mirror and fiber input facet does not bring any aberrations, tilts or translations with respect to the fiber center, one can compute the amplitude of the modes that were launched into the fiber. Such a mode expansion leads to the mode power distributions illustrated in Fig.6, for obtaining optimal self-cleaning on the $LP_{01}$ (Fig.6-a), $LP_{11}$ (Fig.6-b), $LP_{12}$ (Fig.6-c) and $LP_{21}$ modes (Fig.6-d), respectively. Fig.6 shows that a broad modal population is always present at the fiber input.

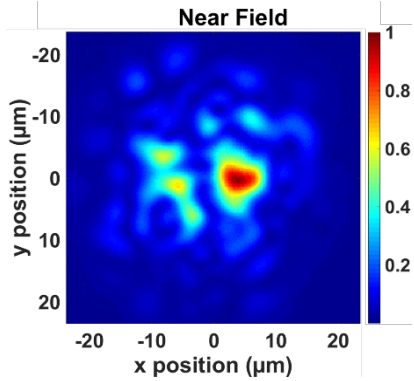

Fig. 7. Near field observed at a relatively low power (150 W), while keeping the input wavefront optimized for self-cleaning on a $LP_{11}$ intensity pattern at high powers (e.g., see Fig 4 above for reference)

Moreover, the mode which prevails (via self-cleaning) after the nonlinear propagation through the MMF already carries a relatively high power fraction (6 to 12%) with respect to the rest of the excited population at the MMF input. However, note the unexpected result that optimized self-cleaning of the $LP_{01}$ mode only requires about 7% of power in that mode at the input, with nearly equal power contribution from the $LP_{31}$ mode. In all final (i.e., after the optimization process was terminated) states of observation, beam propagation in the MMF remains well in the nonlinear regime.

A first signature of the nonlinear propagation regime for optimized many-mode self-cleaning can be found in the output spectra, where strong spectral broadening (~x30) is observed around the input laser line of 0.3 nm width at - 10dB. A second signature of the impact of nonlinearity was visible, when we imposed a sudden decrease of the launched power. Keeping the optimized profiling of the input to achieve Kerr self-cleaning on a given target mode, i.e., maintaining the same input conditions, beam shaping at the fiber output completely vanished when the input power was reduced below the threshold for nonlinear effects. A typical case for destruction of a $LP_{11}$ self-cleaning as shown in Fig.4 is given in Fig.7: a single bright spot remains, sitting on a background of speckles. Our results clearly show that selection of a LOM at the fiber output end did not trivially result from the tailored pure excitation of the same mode at input. An experiment further confirms this statement. The tentative linear selection of a LOM, following the same procedure, failed at low powers. At an input power of 160 W, *i.e.*, in the linear propagation regime, we tried to select a pure LOM at the fiber output following the same procedure as above. The case of a $LP_{11}$ mode is illustrated in Fig.8. Optimized beam shaping at the input, gave an output image having a high correlation with the $LP_{11}$ intensity distribution, hence minimizing the cost function.

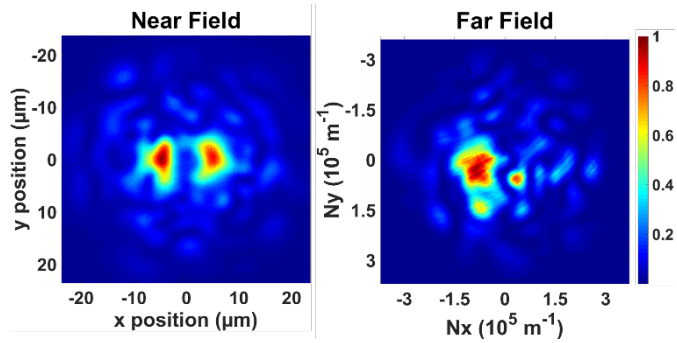

Fig. 8. Near field and far field images recorded after optimization of the input wavefront to get a $LP_{11}$ intensity pattern at the MMF output in linear propagation regime. Although the figure on the right resembles a $LP_{11}$ (with a strong background) the far field indicates $LP_{11}$ is not the major modal contribution to the output field.

However, in this case the output far field did not evolve into that of the desired target mode, contrary to what happened in the nonlinear self-cleaning regime. Therefore, from our recorded data it is not possible to conclude that light delivered by the MMF after optimization in the linear regime has most of its power carried by the desired target mode (see Appendix). When operating in the linear regime, a cost function for optimization, based on intensity correlation in the near field only, is not sufficient to channel guided light into a desired mode. This is due to the fact that, in the linear regime, there exist many options for modal interferences that may lead to any given intensity pattern at the MMF output (see Appendix). To the contrary, with nonlinear mode coupling induced by Kerr effect, the various mode field amplitudes and phases are connected to each other, so that solutions of the same inverse problem might be scarce. As a result, the optimization process results into an output field with always a major contribution provided by the desired target mode. This is one of the major conclusion of our investigations. Both the excitation of an appropriate input modal population, and a sufficiently high power, were necessary to trigger nonlinear dynamics which favored power channeling into a specific LOM during propagation.

Lastly, in order to show that the fiber output results from nonlinear evolution towards a steady-state with a nearly pure mode shape, we performed a fiber cut-back experiment (see Fig. 9).

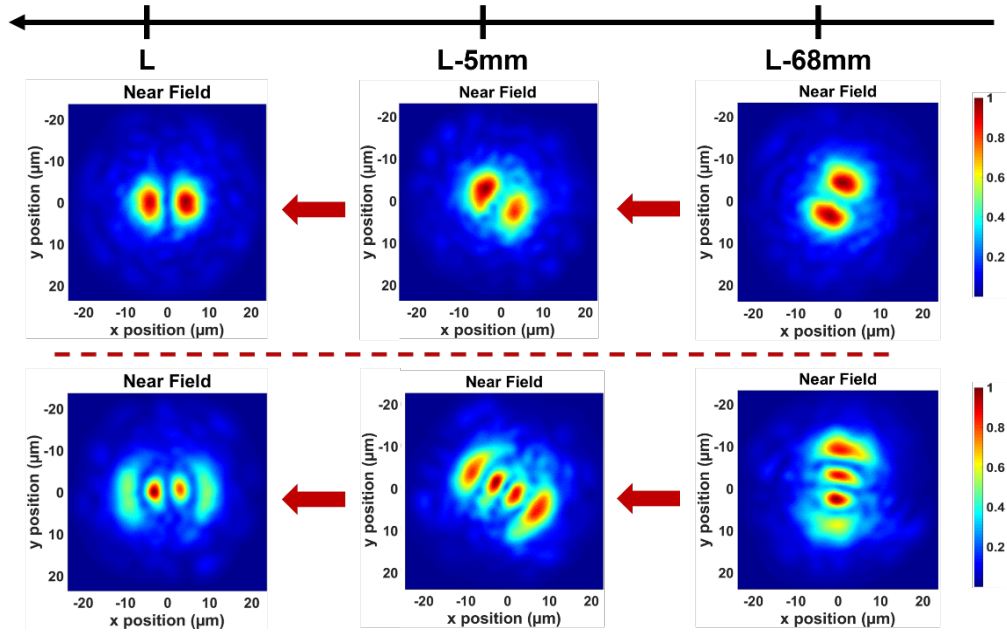

Fig. 9. Near-field images after optimized Kerr self-cleaning of two examples of LOMs ($LP_{11}$ top row, $LP_{12}$ bottom row) recorded at the fiber output (left column), then after cut back of the fiber by 5 mm whilst keeping the whole input conditions (center column), and after cut back by 68 mm (right column). Translational invariance of the cleaned beam was observed. Rotation of the patterns in the camera frame from left to right are due to fiber twist.

After optimization at high power of the input laser profile, in order to obtain a given output mode profile by means of Kerr self-cleaning, the fiber end was removed from the holder and cleaved, in order to progressively reduce the fiber length. The fiber was then re-positioned onto the holder, and the delivered light beam was imaged on a camera for comparison with the previous pattern. We observed that the guided wave pattern was almost the same at either 5 mm or at 68 mm away in terms of propagation distance from the initial fiber output (see Fig.9). This is very unlikely to occur in the linear propagation regime, where the phase relationships between the modes should rapidly vary with the propagation distance, which quickly modifies their interference pattern (except at some specific positions of self-imaging). Along the same line, in the initial state of optimized Kerr self-cleaning on a LOM, we gently perturbed the fiber by hand, by introducing a displacement or bending. The output pattern did not change much under these perturbations, showing that the robustness of self-cleaned beam also applies to LOMs, although to a lesser extent than for the fundamental mode.

## 4. Conclusions

Nonlinear coupling of optical waves guided in a multimode optical fiber can be mastered by a fine tuning of the transverse profile of the laser excitation. We demonstrated here that peculiar spatial nonlinear dynamics, namely, the effect of Kerr beam self-cleaning, can be adaptively controlled in order to shape at high powers a stable beam delivered by the MMF on a variety of patterns, different from the single bell structure of the fundamental mode. A phase pattern, imprinted on the laser beam that is launched into the MMF by reflection on a deformable mirror, served for the control of the multimode propagation, and it was iteratively adjusted in order to obtain a specific output intensity profile. Kerr self-cleaning on various LOMs ($LP_{11}$, $LP_{02}$, $LP_{12}$, $LP_{21}$ and $LP_{22}$) was observed for the first time, thanks to the extra degrees of freedom brought by wavefront tailoring. For all states of spatial beam self-organization, a broad modal population was excited, and the mode expected to prevail after nonlinear propagation counts among the ones with the highest power fraction in the excited population. Comparison between observations in the linear and the nonlinear propagation regimes proved that genuine nonlinear evolution assisted by the Kerr effect was involved in the selection of a desired mode at output. The lowest power threshold for self-cleaning is that for the fundamental mode. All other modes studied in this work, although of low-order, exhibited a significantly higher threshold for self-cleaning. Cut-back experiments have shown that Kerr-cleaned LOM beams propagate unchanged, at least towards the output end of the fiber.

Our results show that marrying the technique of wavefront shaping with nonlinear transverse mode coupling leads to many-mode Kerr beam self-cleaning. That is, we could generate a whole stable alphabet of pure fiber modes that propagate independently and nearly unchanged along a MMF, irrespective of random linear mode coupling or fiber bending. We anticipate that this finding will have important implications for a number of applications ranging from optical communications using spatial division multiplexing, nonlinear microscopy and endoscopy, and micromachining.

**Funding**

The European Research Council (ERC) under the European Union's Horizon 2020 research and innovation programme (grant No. 740355); The French Agence Nationale de la Recherche, POMAD project( grant N°ANR-14-CE26-0035-01); The French "Investissements d'Avenir" program, project ISITE-BFC (grant N°ANR-15-IDEX-0003); CILAS- Ariane Group (grant N°0C1XBB007).

**Appendix**

In this section, we show that the near-field intensity distribution is not sufficient to prove the major presence of a given mode in the field which is delivered by a GRIN MMF. The phase transverse distribution is crucial on that matter. In principle, the phase distribution can be completely recovered by measurements, such as for example from its interference with a plane wave reference. But a sufficient signature of the phase can be found in the far-field intensity pattern, which is proportional to the squared amplitude of the Fourier transform of the near-field. It is known that some phase retrieval algorithms are even based on the sole knowledge of the near-field and far-field intensity images [25]. In the particular case of a GRIN fiber, its LP mode fields are described by Laguerre-Gaussian profiles which are self-invariant upon a Fourier transform. Therefore, by looking at both the image from the fiber output facet, and the image after diffraction into the far-field, should be sufficient to decide whether or not there is a major contribution of a particular mode to the output beam. To illustrate this point, we considered two optical fields composed by a linear combination of guided modes of the GRIN fiber of the present work, in order to reproduce situations that could be observed at the output of the fiber, as a result of both the input launching conditions, and fiber stress and bending along the fiber. In the first case, we considered a pure $LP_{11}$ mode, where the two lobes are out of phase. In top left and center panels of figure A1, we show the image of the near-field, as well as the corresponding image in the far-field. As expected, the far-field image shows a similar pattern to that in the near-field, since the $LP_{11}$ mode is an eigenfunction of the Fourier transform.

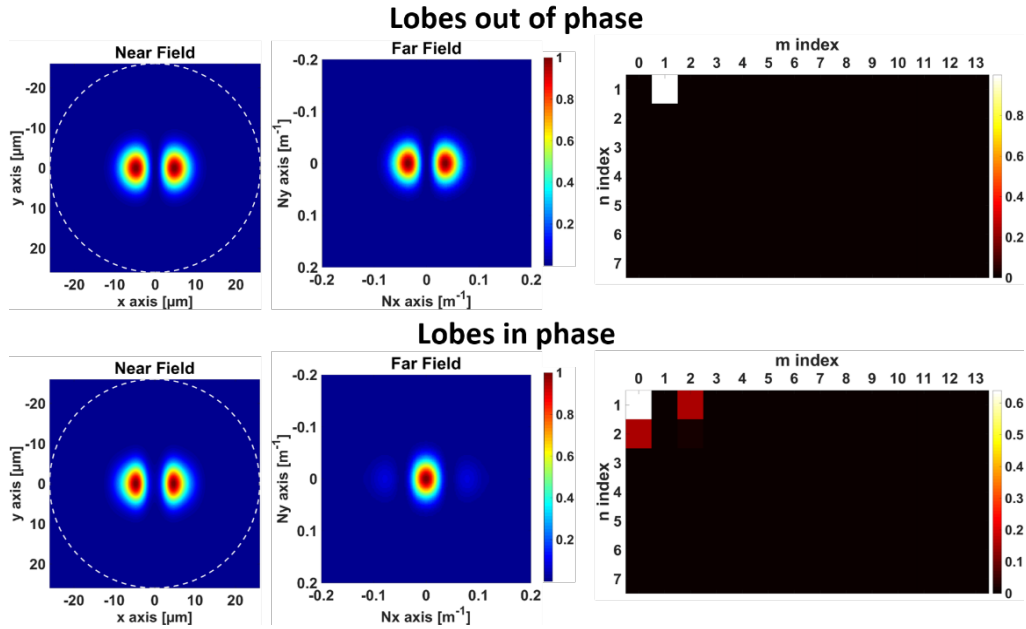

Fig. A1. Comparison of two situations with an almost similar intensity pattern in the fiber output near-field (left column), but different modal content: the case of a single $LP_{11}$ mode (top row) is compared with a mixture of $LP_{01}$, $LP_{02}$, and $LP_{21}$ modes (bottom row). The signature of the different modal contents (right column) is well visible in the far field (center column).

The top right panel in Fig.A1 shows the mode map, which confirms the genuine presence of a single mode. In the bottom row of Fig.A1, we illustrate the results of a second case, where the two field lobes are in phase. As we can see, the near-field image can be easily confused with that of an $LP_{11}$ mode, especially in the presence of a speckled background as it is often the case in the experiments. However, the corresponding far-field image reveals the information that the lobes are in phase. The corresponding mode map (bottom right panel in Fig. A1) shows that the modal composition is a mix of even-parity modes, as expected by the symmetry of the field profile. We also considered other cases, in which we applied a scaling factor to modify the distance and the size of lobes, and always obtained similar results (here not shown). From these results, we may conclude that the near-field intensity profile is not sufficient to characterize modes such as the $LP_{11}$ mode. This also validates our approach, based on the additional monitoring of far-field images, which permits to unveil additional key information that is important for the identification of pure MMF modes.

## References

1. I. M Vellekoop, and A. P Mosk, "Focusing coherent light through opaque strongly scattering media," Opt. Lett. **32**, 16, 2309–2311 (2007).
2. S. Popoff, G. Lerosey, R. Carminati, M. Fink, A. C. Boccara, and S. Gigan, "Measuring the transmission matrix in optics: an approach to the study and control of light propagation in disordered media," Phys. Rev. Lett. 104 (10), 100601 (2010).
3. S. Popoff, G. Lerosey, M. Fink, A. C. Boccara, and S. Gigan, "Image transmission through an opaque material " Nat. Commun **1**, 81 (2010). DOI: 10.1038/ncomms1078
4. E. Small, O. Katz, Y. Bromberg and Y. Silberberg, "Focusing and compression of ultrashort pulses through scattering media," Nat. Photonics 5, pp. 372–377 (2011)
5. M. Kim, Y. Choi, C. Yoon, W. Choi, J. Kim, Q-H. Park and W. Choi, "Maximal energy transport through disordered media with the implementation of transmission Eigenchannels," Nat. Photonics, 6, pp. 581–585 (2012).
6. J. Aulbach, B. Gjonaj, P. M. Johnson, A. P. Mosk, and A. Lagendijk, "Control of Light Transmission through Opaque Scattering Media in Space and Time," Phys. Rev. Lett. 106, 103901 (2011)

7. H. Frostig, E. Small, A. Daniel, P. Oulevey, S. Derevyanko, and Y. Silberberg, "Focusing light by wavefront shaping through disorder and nonlinearity," Optica **4**, 1073-1079 (2017).
8. S. Rotter and S. Gigan, "Light fields in complex media: Mesoscopic scattering meets wave control," Rev. Mod. Phys. **89**, 015005 (2017)
9. T. Čižmár and K. Dholakia, "Exploiting multimode waveguides for pure fibre-based imaging," Nat. Commun. **3**, 1027 (2012).
10. J. Carpenter, B. J. Eggleton, and J. Schröder, "110x110 optical mode transfer matrix inversion," Opt. Express **22**, 96-101 (2014).
11. D. Loterie, S. Farahi, I. Papadopoulos, A. Goy, D. Psaltis, and C. Moser, "Digital confocal microscopy through a multimode fiber," Opt. Express **23**, 23845-23858 (2015).
12. R. Florentin, V. Kermene, J. Benoist, A. Desfarges-Berthelemot, D. Pagnoux, A. Barthelemy and J-P Huignard, "Shaping the light amplified in a multimode fiber," Light: Science and Applications; **6**, 2 :e16208 (2017).
13. W. H. Renninger and F. W. Wise, "Optical solitons in graded-index multimode fibres," Nat. Commun. 4, 1719 (2013).
14. K. Krupa, A. Tonello, A. Barthélémy, V. Couderc, B. M. Shalaby, A. Bendahmane, G. Millot, and S. Wabnitz, "Observation of Geometric Parametric Instability Induced by the Periodic Spatial Self-Imaging of Multimode Waves," Phys. Rev. Lett. 116, 183901 (2016).
15. Z. Liu, L. G. Wright, D. N. Christodoulides, and F. W. Wise, "Kerr self-cleaning of femtosecond-pulsed beams in graded-index multimode fiber," Opt. Lett. 41, 3675-3678 (2016).
16. L. G. Wright, D. N. Christodoulides, and F. W. Wise, "Controllable spatiotemporal nonlinear effects in multimode fibres," Nat. Photon. 9, 306 (2015).
17. K.Krupa, A Tonello, B. Shalaby, M. Fabert, A. Barthélémy, G. Millot, S. Wabnitz, V. Couderc, "Spatial beam self-cleaning in multimode fiber," Nat. Photonics 11 (4), 237-241, (2017)
18. G. Lopez-Galmiche, Z. Sanjabi Eznaveh, M. A. Eftekhar, J. Antonio Lopez, L. G. Wright, F. Wise, D. Christodoulides, and R. Amezcua Correa, "Visible supercontinuum generation in a graded index multimode fiber pumped at 1064nm," Opt. Lett. 41 (11), 2553–2556 (2016).
19. K. Krupa, C. Louot, V. Couderc, M. Fabert, R. Guenard, B. M. Shalaby, A. Tonello, A. Barthélémy, D. Pagnoux, P. Leproux, A. Bendahmane, R. Dupiol, G. Millot, S. Wabnitz, "Spatiotemporal Characterization of Supercontinuum Extending from the Visible to the Mid-Infrared in Multimode Graded-Index Optical Fiber," Opt. Lett. 41, 5785-5788 (2016).
20. O. Tzang, A. M. Caravaca-Aguirre, K. Wagner, and R. Piestun, "Adaptive wavefront shaping for controlling nonlinear multimode interactions in optical fibres," Nat. Photonics 12, 368–374 (2018).
21. M. A. Eftekhar, L. G. Wright, M. S. Mills, M. Kolesik, R. Amezcua Correa, F. W. Wise, and D. N. Christodoulides, "Versatile supercontinuum generation in parabolic multimode optical fibers," Opt. Express 25, 9078-9087 (2017)
22. Peng Lu, Matthew Shipton, Anbo Wang, Shay Soker, and Yong Xu, "Adaptive control of waveguide modes in a two-mode-fiber," Opt. Express 22, 2955-2964 (2014)
23. I. M. Vellekoop, "Feedback-based wavefront shaping," Opt. Express **23**, 12189-12206 (2015).
24. E. Deliancourt, M. Fabert, A. Tonello, K. Krupa, A. Desfarges-Berthelemot, V. Kermene, G. Millot, A. Barthelemy, S. Wabnitz, and V. Couderc, "Kerr beam self-cleaning on the $LP_{11}$ mode in graded-index multimode fiber" to appear in OSA Continuum (2019).
25. R. W. Gerchberg and W. O. Saxton. "Practical algorithm for determination of phase from image and diffraction plane pictures," Optik, 35(2) :237-246,1972.